\documentclass[twocolumn,showpacs,preprintnumbers,amsmath,amssymb]{revtex4}
\usepackage{graphicx}
\begin{document}
\def\be{\begin{equation}}
\def\ee{\end{equation}}
\def\bearr{\begin{eqnarray}}
\def\eearr{\end{eqnarray}}
\def\tc{$T_c~$}
\def\tcl{$T_c^{1*}~$}
\def\c2{ CuO$_2~$}
\def\ruo{ RuO$_2~$}
\def\lsco{LSCO~}
\def\bi{bI-2201~}
\def\tl{Tl-2201~}
\def\hg{Hg-1201~}
\def\sro{$\rm{Sr_2 Ru O_4}$~}
\def\rc{$RuSr_2Gd Cu_2 O_8$~}
\def\mgb{$MgB_2$~}
\def\pz{$p_z$~}
\def\ppi{$p\pi$~}
\def\sqo{$S(q,\omega)$~}
\def\tperp{$t_{\perp}$~}
\def\he4{${\rm {}^4He}$~}
\def\ags{${\rm Ag_5 Pb_2O_6}$~}
\def\nxcob{$\rm{Na_x CoO_2.yH_2O}$~}
\def\lsco{$\rm{La_{2-x}Sr_xCuO_4}$~}
\def\lco{$\rm{La_2CuO_4}$~}
\def\lbco{$\rm{La_{2-x}Ba_x CuO_4}$~}
\def\half{$\frac{1}{2}$~}
\def\thalf{$\frac{3}{2}$~}
\def\tst{${\rm T^*$~}}
\def\tch{${\rm T_{ch}$~}}
\def\jeff{${\rm J_{eff}$~}}
\def\nbc{${\rm LuNi_2B_2C}$~}
\def\cabc{${\rm CaB_2C_2}$~}
\def\nboo{${\rm NbO_2}$~}
\def\voo{${\rm VO_2}$~}
\def\nip{$\rm LaONiP$~}
\def\nisb{$\rm LaONiSb$~}
\def\nibi{$\rm LaONiBi$~}
\def\fep{$\rm LaOFeP$~}
\def\cop{$\rm LaOCoP$~}
\def\mnp{$\rm LaOMnP$~}
\def\fesb{$\rm LaOFeSb$~}
\def\febi{$\rm LaOFeBi$~}
\def\efeas{$\rm LaO_{1-x}F_xFeAs$~}
\def\hfeas{$\rm La_{1-x}Sr_xOFeAs$~}
\def\hSfeas{$\rm Sm_{1-x}Sr_xOFeAs$~}
\def\hCefeas{$\rm Ce_{1-x}Sr_xOFeAs$~}
\def\feas{$\rm LaOFeAs$~}
\def\Ndfeas{$\rm NdOFeAs$~}
\def\Smfeas{$\rm SmOFeAs$~}
\def\Prfeas{$\rm PrOFeAs$~}
\def\refeas{$\rm REOFeAs$~}
\def\refesb{$\rm REOFeSb$~}
\def\refebi{$\rm REOFeBi$~}
\def\ttog{$\rm t_{2g}$~}
\def\fese{$\rm FeSe$~}
\def\fete{$\rm FeTe$~}
\def\eg{$\rm e_{g}$~}
\def\dxy{$\rm d_{xy}$~}
\def\dzx{$\rm d_{zx}$~}
\def\dzy{$\rm d_{zy}$~}
\def\dxsq{$\rm d_{x^{2}-y^{2}}$~}
\def\dzsq{$\rm d_{z^{2}}$~}
\def\LAO{$\rm LaAlO_3$~}
\def\STO{$\rm SrTiO_3$~}
\def\hsm{$\rm H_2 S$~}
\def\hm{$\rm H_2$~}
\def\silane{$\rm Si H_4$~}

\title{ Emergent Mott Insulators and Possibility of High T$_{\bf c}$ Superconductivity in\\
Pressurized Molecular Solids, H$_2$S, SiH$_4$, H$_2$ etc.}

\author{ G. Baskaran}
\affiliation
{The Institute of Mathematical Sciences, C.I.T. Campus, Chennai 600 113, India \&\\
Perimeter Institute for Theoretical Physics, Waterloo, ON, Canada}

\begin{abstract}
Paired valence electrons forming covalent bonds in molecues are \emph{confined cooper pair correlations}.  High pressure dissociates molecules in solid  \hsm, \silane, \hm etc. and form extended structures. However, valence electrons are resilient and continue to bond and sometimes resonate. It is suggested that some structures contain \emph{emergent Mott insulators} (EMI) and support superconductivity and other phases, under \emph{special conditions.}.  In pressurized solid \hsm, we propose presence of a sublattice of relatively narrow, nearly half filled band of H atoms, trapped in ordered interstitials of colvalently bonded S atom network. EMI offers a high pairing energy scale. A theory for recently claimed superconductivity in solid \hsm (Tc $\sim$ 205 K), Silane (Tc $\sim 17 K$), and systems like \hm is presented.
\end{abstract}

\maketitle

Pressure is a remarkable and unique variable at our disposal. It could create novel states of matter, not easily reachable otherwise.  A resilience of quantum matter results in unexpected strategies and adaptation to high pressures. Quantum mechanics seems to play a deep role here. The rich phase diagram of ice is an example. Hydrogen bonds, a subtle consequence of quantum mechanics, are preserved and their networks modified in imaginative ways to accommodate external pressure.

One observes similar resilience in molecular crystals and other solids under pressure \cite{ReviewHemleyMolecularSolid}. Here covalent bonds, a basic quantum mechanical pairing of valence electrons in spin singlet state,  are resilient. Pressure aapplied to \hm, \hsm, \silane breaks bonds and dissociates molecules. But new extended structures containing a rich variety of covalent bond networks get created, over a range of pressures. A resilience from valence bonds seem to push to higher pressures, the onset predicted by Wigner-Huntington \cite{WignerHuntington} of one electron metallization and onset of room temperature predicted  by Ashcroft  \cite{Ashcroft1968} of room temperature superconductivity. 

In this background, a recent high pressure study by Drozdov, Eremets and Troayan \cite{DET1} make a remarkable claim of strong signals for superconductivity at Tc$\sim$ 190 K , in \hsm at a pressure of 190 GPa. A second study from the same group reports \cite{DET2} an even higher Tc $\sim$ 205 K at 160 GPa.  Once superconductivity is confirmed by Messner effect and zero resistance measurements, this discovery will surpasses earlier and reproducible highest Tc $~\sim$ 164 K in trilayer cuprates \cite{Shilling}. It is exciting and prospects for room temperature superconductivity seem closer. Another hydrogen rich solid, Silane, \silane, similar to \hsm has been also shown to exhibit superconductivity \cite{EremetsSilane}, albeit with a lower Tc $\sim$ 17 K at a lower pressure of 113 GPa.

We suggest that molecular solids offer an interesting route to high Tc superconductivity, using available valence electrons and their strong pairing (chemical bonding) tendencies. We suggest that these singlet bonds are confined Cooper pairs. Based on available informations on structures formed after molecular dissociation, certain packing and theoretical considerations, we suggest that new structures might contain sublattices that are close to a Mott insulator or a doped Mott insulator.  These Emergent Mott Insulators (EMI) are likely abode  for high Tc superconductivity and other novel phases.

Mott insulators could emerge in \hsm and \silane, \emph{under some restrictive conditions} as follows. After pressure induced dissociation, S or Si atoms form their own extended saturated covalent bond network and satisfy (saturate) desires of their own valence electrons to form bonds. Molecular hydrogen and atomic hydrogen get accommodated in suitable interstitials, to maximize cohesion with the big atoms as well as among themselves.  

When neutral H atoms get  accommodated in interstitials in an ordered fashion a desired situation arises. As interstitial separations of the big S or  Si atoms are are large compared to H-H distaince in \hm molecule, atomic hydrogen subsystem forms an expanded lattice. However, 1s orbitals of H-atoms hybridize through S or Si orbitals and narrow band physics could emerge.

Conditions for Mott physics to emerge is i) comparable band widths and coulomb energy or Hubbard U,  ii) an apropriate electronegativity difference between S, Si systems and H subsystems, to get an optimal charge transfer resulting in a doped Mott insulator and iii) avoidance of strongly dimerized H atom sublattice.

Pressure as a variable will not fine tune systems to desired Mott phases capable of supporting high Tc superconducting phase, for example. Further pressure can have desired effects only in a limited pressure range. Beyond a pressure of $\sim$ 200 GPa, as seen in a variety of band structure calculations, valence and conduction bands become too broad to have significant electron correlation effects.

Spin-half Mott insulators have antiferromagnetic exchange interacions. Spin singlet valence bonds are formed and resonate as unsaturated covalent bonds or manifest as long range antiferromagnetic order.  In a linear chain valence bonds resonate. A dimerization in the chain will reduce quantum mechanical resonance and encourage valence bond localization. Similar physics takes place in 2 dimensional and 3 dimensional lattices.  

According to RVB theory \cite{PWAScience,BZA,BAGauge,ABZH,GBIran} valence bonds or paired valence electrons are \emph{preexisting and delocalized neutral cooper pairs}. Doping leads to charged valence bonds resulting in superconductivity. Scale of superconducting Tc, for the layered cuprates is \cite{ABZH} k$_B T_c  \sim  t \delta $, where $\delta$ is the dopant density and t, a hopping energy scale. If an interchain pair tunneling causes superconductivity \cite{WHA} k$_B T_{c} \sim \frac{4 t^2_{\perp}}{J}$, where J is a spin gap scale and t$_\perp$ is the interchain hoping matrix element. 

A body of theoretical works on superconductivity in solid \hsm \cite{Prediction} and solid H$_3$S \cite{H3SPrediction}, a prediction before the experimt and theories that followed the experiments \cite{Hirsch, Picket,Mazin,Arita,Duan,Errea,Flores} exists.  Our mechanism is different. We find a compelling phenomenology, and certain microscopic arguments in favor of a strong correlation based mechanism and interesting Mott physics, in an intermediate pressure range, before one electron metallization occurs.

In what follows we present our proposal for superconductivity for Silane, \hsm and suggest microscopic models and end with a discussion on solid \hm that has been of great interest to the physics community. Hydrogen rich solids, including silane, was suggested to be a good route to achieve he light hydrogen atom based high Tc superconductivity by Ashcroft \cite{SilaneAshcroft}

In Pressurized solid \silane \cite{EremetsSilane} molecular dissociation occurs at $\sim$ 50 GPa; aximum superconducting Tc $\sim$ 17 K at 113 GPa. Doubts have been raised about the claim of bulk superconductivity in silane.  Such a superconductivity is not impossible, from theory point of view \cite{SilaneAshcroft}

In the XRay studies \cite{EremetsSilane} of the superconducting sample, Si atoms form a close packed hcp lattice. Lighter H atoms could not be located.  The authors observe that  i) Si atoms with a covalent radius $\sim$ 1.117 \AA  ~ are close packed, ii) H atoms with a covalent radius $\sim$ 0.37 \AA ~  could just fit in the small tetrahedral interstitial voids and iii) two H atom could fit in the octahedral interstitial space.

Close packing and a right covalent radii, make Si and H atoms nearly neutral. Covalency dominates. Further, in strained conditions atoms like Si can make non tetrahedral bonds, by hybridization with other available empty orbitals. In a first approximation, Si atoms suitably bond, have saturated covalent bonds; in a band picture we have filled bands. Inerstitials of two face sharing tetrahedra of Si atoms trap two H atoms and isolaes them electronically. We assume that these H atom pairs get bonded and are below the chemical potential. This takes care of two H atoms per formula unit \silane in our attempt to model low energy physics of electrons.

In the hcp structure Si octahedra share faces and run along c-axis as chains. These chains form a triangular lattice array. Collection of H atoms trapped in octahedral interstitials, in the first approximation, form an uninterrupted and uniform chain atoms along c-axis.  Symmetry (two atoms in the tetrahedral interstitial) allows dimerization of H atoms along the chain. Mean nearest neighbor H-H distance along the chain is $\sim 1.18 \AA$. Nearest neighbor H-atom chain separation is  $\sim 2.67 \AA $. So we have H atom chains with weak interchain coupling.

Overlap of 1s wave function of nearest neighbor H atoms along the chains is small, inview of an expanded H-H distance. However, they strongly hybridize through Si orbitals and form a quasi one dimensional tight binding and nearly half filled bands. Our minimal model is a coupled uniform one dimensional chain Hubard model, with (inter, intra) chain nearest neighbor hopping parameter (t, t$_\perp$) and Hubbard U.
\bearr
H &=&  -t \sum_{i m} c^\dagger_{i m\sigma} c_{i+1 m\sigma} + H.c. + U \sum_i n_{i m\uparrow} n_{i m\downarrow} + \nonumber \\
  &+& - t_{\perp} \sum_{\langle m n \rangle} c^\dagger_{i m\sigma} c_{i  n\sigma} + H.c.
\eearr
In the site index (i,m) i denotes position along the chain and m the ab-plane coordinate of the chain. Using band structure results available for solid \silane, H-rich solids and some quantum chemistry we make a rough estimate of the parameters:   t $\sim$ 1 to 2 eV, t' $\sim$ 0.1 eV and U $\sim$ 5 to 10 eV.  In the above idealized approximation, we have a Mott insulator at Half filling, since U$\gg$ t, In turn we have weakly coupled spin-\half Heisenberg chains. 

Doping needed for superconductivity could be obtained from \emph{self doping}, a charge transfer between Si and H subsystems,  as their electronegativity are different in general. In a band picture, broad filled valence bands or empty conduction bands, primarily of Si character cross the chemical potential and transfer charges to the Mott insulating H-subsystem. 

In the RVB theory the above model supports singlet superconductivity with a dome in the Tc versus doping axis. Individual doped H-atom chain have a strong pairing correlations, but has divergent quantum and thermal fluctuations. Three dimensional superconductivity arises from coherent interchain pair tunneling. In this mechanism \cite{WHA}, $k_B T_c  \sim \frac{t^2_{\perp}}{J}$. Here J $\sim \frac{4t^2}{U}$ is the spin gap scale in the H atom chain. This gives us an upper bound, or an optimistical scale of Tc $\sim$ 100 K, at optimal doping.

Pressure generated structures are expeced to be fragile in general, as several balancing forces, quantum, solid state chemistry, screening etc. are at work. Doping need not be a linear function of pressure. Pressure vs Tc curve in the only available experiment seems to indicate, interpreted in the light of EMI,  a rather fast change in internal doping in a narrow pressure range, in the background of issues like phase separation, competing instabilities etc. Uniform chain is an idealization and dimerization will reduce valence bond resonance and will reduce Tc.

Superconducting Tc, symmetry of the order parameter etc., become a quantitative and difficult issues requiring more phenomenological and theoretical inputs, not available at the moment.  Our main message is to recognizing possibility of emergence of spin-half Mott insulating H atom subsystem in a narrow range of pressure that prepares grounds for high Tc superconductivity. With the above considerations we will move on to pressurized solid \hsm, where the claimed Tc is impressively high $\sim$ 190 K, but much less phenomeoogy is known

Superconductivity in solid \hsm \cite{DET1,DET2} in the form of strong signals, with a maximum Tc $\sim$ 190 K and 203 K, at a pressurez in the range $\sim$ 180 GPa, has been reported. A sharp resistivity drop to significantly low but nonzero value and a magnetic field dependence of Tc, expected of superconductors is seen.  An isotope effect on Tc, is also seen by deuterium substitution.  No structural information is available.  However theoretically predicted structures are available. Further theories predict formation of a new structure H$_3$S \cite{H3SPrediction,H3SHemley}, in a phase separation scenario, in pressurized \hsm.

We try to use some insights gained from \silane. Si and S atoms have different quantum chemistry. However, covalent bond radii of Si and S are are close, 1.117 and 1.04 \AA  , respectively. Sulfur, like their cousins Se and Te, is known to create two saturated valence bonds and form helical chains.  Earlier experimental results \cite{JapaneseExptDissociationStructure2004}and theories\cite{Parrinello2000} suggest that after molecular dissociation, i) packed helical chains of sulfur atoms are formed and ii) H atoms fill the interstitial spaces. Further, earlier theoretical simulation \cite{Parrinello2000} works have suggested formation of stacked sheets of sulfur atoms and hydrogen atoms.  In addition, in a molecular dissociation and phase separation scenario, a more stable stoichiometric phase H$_3$S, has been suggested and widely analysed theoretically for superconductivity. In the recent theoretically determined structures, for both \hsm and H$_3$S, H atoms selectively occupy \emph{bond centered interstitial sites}. 

We suggest that dissociated H atoms fill (bond cenered or other) interstitial voids as \hm molecule or H atoms and creates opportunity to form Mott insulating H atom susbsystem.  We get support for this from an available band structure result \cite{Prediction} for \hsm in the P-1 structure, predicted to be stable around 130 GPa. It supports a nearly half filled and nearly isolated band at the fermi level. This band has an admixture of H and S atom orbitals. Width of this band is close to 4 eV, smaller than the Hubbard U for H-atom $\sim$ 8 eV.  This band arises from H-atom chains occupying bond centered interstitials in the P-1 structure.  Valence electron localization function (ELF) indicates strong hybridization of H orbital to sulfur orbitals. There are filled valence bonds just crossing the fermi level, indicating a differing electronegativity between the S and H subsystem and possibility of internal charge transfer and doping of Mott insulator. 
 
We have also looked other possibilities like sheets of H atoms intercalated between Sulfur layers, a structure suggested by earlier simulation studies. We will not go into details. We have fascinating possibility of doped Mott insulator and chiral superconductivity, in the case of triangular or honeycomb H sheets. The Tc's in this case are more favorable, reaching 200-300 K scale. We believe that emergence of Mott insulator is similar to \silane, but with different parameters.  In the present case, for the P-1 structure we have weakly coupled Mott insulating chains (equation 1). For the range of our estimates of t, t$_\perp$ and U for \hsm we get a Tc in the range of 100  K. We wish to emphasize that what is importantly is our suggestion of emergent Mott insulator, seat of high Tc superconductivity.  Now we go on to discuss solid \hm.

Superconductivity in Solid \hm is long awaited, after the important theoretical suggestion of Ashcroft in 1968. Understanding the nature of high pressure phases of solid \hm is very important for planetary physics.  Metallization studies continue both experimental \cite{MetallicHydrogenEremets} and theoretical fronts.  Extensive theoretical studies provide excellent insights \cite{AshcroftHoffman,DissociationH2} about why metallization gets delayed to much higher values of pressure than predicted value of 25 GPa by Wigner and Huntington. 

An exciting experiment in 2012 \cite{ExptGrapheneLikeHSolid} indicated presence of graphene like hexagonal layers of H atoms trapping layers of \hm molecules around 200 GPa. This started many theoretical activities suggesting possible structures \cite{HemleyH2Graphene}. In a recent work Nuomov and Hemley \cite{HemleyH2Graphene1}associate this tendency to aromaticity and closed effects known in benzene and p-$\pi$ bonded systens. What is exciting to us is a selfconsistent formation two subsystems in a single element system, i)  a molecular lattice containing \hm molecule or benzene like H$_6$ molecules and ii) a coexisting atomic lattice of H atoms. In the molecular subsystems valence electrons are paired as valence bonds and localized.

Most structures available in the literature, for the 200 to 300 GPa pressure range exhibit a band gap. Valence and conduction bands widths are $\sim$ 10 eV.
In some cases gaps seem to arise from (Kekule) dimerized two dimensional structures or dimerized chains. We model, as a starting point, the system as weakly coupled layers of honeycomb lattice Hubbard models. The overall band width is large $\sim$ 20 eV. Since neutral \hm molecular system is less polarizable, we expect less screening and a Hubbard U $\sim$ 10 eV or more.

Since bandwidth exceeds Hubbard U, Mott localization will be absent in the sheets of H atoms. However, 2 dimensionality enables electron correlation based high Tc superconductivity, even for intermediate value of U. This has been discussed at some length in the theory of high Tc superconductivity in doped graphene \cite{PathakShenoyGB} and very recently for silicene \cite{GBSilicene}. If we use theoretical results available for superconductivity in graphene, and scale it to the case of hydrogen honeycomb lattice we again get encouragingly high Tc $\sim$ 200 K and possibility of unconventional pairing symmetries.

To conclude, pressure modified landscape of crystal structures is rich. In the present article we have suggested presence of extended substructures, where electron correlations become important, resulting in possibilities such as high Tc superconductivity. Strongly overlaping s and p orbitals provide a large bandwidth. At the same time the tightly bound valence electron in H atom provides a possibility for very high U = Ionization energy - Electron affinity $\sim$ 13 eV.
So we seem to have an advantage over cuprates by having similar physics but with scaledup parameters. Cuprate family, with a band width is in the range $\sim$ 2 to 3 eV, has already achieved superconductivity with a Tc as high as 163 K in trilayer cuprates. May be we can expect more from EMI's.

The path to high Tc superconductivity is paved with serious hurdles, a lesson we learn from cuprates. Competing orders very often degrade superconducting Tc.  Unscreened coulomb interactions, electron-lattice interactions and disorder encourage competing orders such as spin stripes, charge stripes and polaronic trapping. In the single layer Bi-2201 cuprate at optimal doping,  a potentially high Tc superconductivity is reduced to a low values of 2 K by competing orders (nobody wins).

An isotope effect seen in the experiments on \hsm can be explained, using a mechanism invoked for cuprates \cite{BhattFisher} through a significant change in amplitude of zero point vibration of the confined H atoms on deuterium substitution.  This change modifies hopping parameter and reduces Tc in the RVB theory of superconductivity. Such an electronic isotope mechanism has been used to explain oxygen isotope substitution effects in cuprates.

In view of several hurdles in the pressure route, favorable situations with a large Tc's will be less likely.  However, a more likely and generic possibility is  a strong local superconducting pairing at high temperatures, as a pseudo gap phase. We encourage more experiments in the pressure range around 100 GPa in hydrogen rich sysems and look for EMIs and local pairing tendencies. It might help sharpen the search for real high Tc superconductivity.  To site an example, in the only work available \cite{EremetsSilane} for \silane, Tc vs pressure curve, when extrapolated, instead of a tranditional `domw' has an intriguing `divergent' tendency to higher values of Tc. A closer experimental scan will be valuable.
s
There are many hydrogen rich solids, molecular solids where we find a role Emergent Mott Insulators. We hope to address  some of them in a future publication.

\textbf{Acknowledgement}.  It is a pleasure to acknowledge V.P.S Awana and Mukul Laad for bringing to my attention reference [4]. I thank Mukul Laad for discussions and insisting on the  importance of RVB physics for superconductivity in \hsm. I thank Science and Engineering Research Board (SERB, India) for a National Fellowship. This work is supported by the Government of Canada through Industry Canada and by the Province of Ontario through the Ministry of Research and Innovation.

\end{document}